\documentclass[letterpaper,twocolumn,10pt]{article}
\makeatletter
\usepackage{mathptmx}  

\usepackage[T1]{fontenc}
\usepackage[utf8]{inputenc}
\usepackage{pslatex}

\usepackage[kerning,spacing]{microtype} 
\usepackage{flushend} 

\usepackage{cite}               
\usepackage{breakurl}           
\usepackage{url}                
\usepackage{xcolor}             
\usepackage[]{hyperref}         
\hypersetup{                    
  colorlinks,
  linkcolor={green!80!black},
  citecolor={red!70!black},
  urlcolor={blue!70!black}
}

\def\maketitle{\par
 \begingroup
   \renewcommand\thefootnote{\fnsymbol{footnote}}%
   \def\@makefnmark{\hbox to\z@{$\m@th^{\@thefnmark}$\hss}}%
    \long\def\@makefntext##1{\parindent 1em\noindent
            \hbox to1.8em{\hss$\m@th^{\@thefnmark}$}##1}%
   \if@twocolumn
     \twocolumn[\@maketitle]%
     \else \newpage
     \global\@topnum\z@
     \@maketitle \fi\@thanks
 \endgroup
 \setcounter{footnote}{0}%
 \let\maketitle\relax
 \let\@maketitle\relax
 \gdef\@thanks{}\gdef\@author{}\gdef\@title{}\let\thanks\relax}

\def\@maketitle{\newpage
 \vbox to 2.5in{
 \vspace*{\fill}
 \vskip 2em
 \begin{center}%
  {\Large\bf \@title \par}%
  \vskip 0.375in minus 0.300in
  {\large\it
   \lineskip .5em
   \begin{tabular}[t]{c}\@author
   \end{tabular}\par}%
 \end{center}%
 \par
 \vspace*{\fill}
 }
}

\def\abstract{\begin{center}%
{\large\bf \abstractname\vspace{-.5em}\vspace{\z@}}%
\end{center}}

\def\section{\@startsection {section}{1}{\z@}{-3.5ex plus-1ex minus
    -.2ex}{2.3ex plus.2ex}{\reset@font\large\bf}}
\makeatother
\usepackage{amsmath,amssymb}
\usepackage{graphicx}
\usepackage{booktabs}
\usepackage{placeins}
\usepackage[bottom]{footmisc}
\usepackage{footnotehyper}

\hypersetup{
  hidelinks,
  pdftitle={Errpedia: Is Your Exact k-NN Vector Search Really Exact?}
}

\title{\Large\bfseries Errpedia: Is Your Exact $k$-NN Vector Search Really Exact?}
\author{%
  \normalfont\normalsize
  \begin{tabular}{@{}c@{\hspace{0.25\textwidth}}c@{}}
    {\large Yinuo Deng} & {\large Ying Zhang} \\
    University of New South Wales & University of Technology Sydney \\
    Sydney, Australia & Sydney, Australia \\
    \href{mailto:yinuo.deng1@unsw.edu.au}{yinuo.deng1@unsw.edu.au} &
    \href{mailto:ying.zhang@uts.edu.au}{ying.zhang@uts.edu.au} \\[0.8em]
    {\large Hanchen Wang} & {\large Wenjie Zhang} \\
    University of Technology Sydney & University of New South Wales \\
    Sydney, Australia & Sydney, Australia \\
    \href{mailto:hanchen.wang@uts.edu.au}{hanchen.wang@uts.edu.au} &
    \href{mailto:wenjie.zhang@unsw.edu.au}{wenjie.zhang@unsw.edu.au}
  \end{tabular}%
}
\date{}

\begin{document}
\maketitle
\pagestyle{plain}
\raggedbottom

\begin{abstract}
Exact $k$-nearest-neighbor ($k$-NN) vector search is a common feature of vector libraries and database systems. It serves applications that require exact results and provides the ground truth for evaluating approximate search. However, floating-point rounding and overflow can cause an ``exact'' search to return mathematically incorrect top-$k$ result sets. These errors can go unnoticed and affect downstream tasks that rely on exact results. We present Errpedia, a purpose-built dataset for testing the numerical correctness of exact vector search. Errpedia contains 1,616 queries across 21 error categories, with ground truth independently verified using exact rational arithmetic. We evaluate 20 widely used library and database configurations and find that every tested configuration returns mathematically incorrect top-10 result sets on at least 1 applicable query. Failures include misranking nearly equidistant vectors and returning incomplete results: 1 library omits all 10 exact matches to a query, and several systems return fewer than $k$ neighbors. We release Errpedia as a reproducible benchmark for developers and researchers to test their implementations. The dataset is publicly available at \url{https://github.com/dynos01/errpedia}.
\end{abstract}

\section{Introduction}
\label{sec:introduction}

Exact $k$-nearest-neighbor ($k$-NN) vector search returns the $k$ vectors closest to a query under a specified distance measure, and is widely supported by vector libraries and database systems~\cite{11202651,10.14778/3685800.3685805}. It serves recommendation workloads~\cite{10.1145/3035918.3064009} and provides ground truth for evaluating approximate nearest neighbor (ANN) search~\cite{AUMULLER2020101374}, training query cost estimators~\cite{10.14778/3836663.3836728}, and learning query termination policies~\cite{10.1145/3749160}. Errors in these answers can therefore affect retrieval, evaluation, and training. Despite its importance as a reference oracle, the numerical correctness of exact vector search implementations is rarely evaluated independently. In particular, existing vector search benchmarks typically focus on the accuracy and performance of approximate search~\cite{10.14778/3718057.3718078,10.1145/3709693}, while the correctness of the underlying exact search is often taken for granted.

Algorithmic exactness does not ensure numerical correctness in floating-point arithmetic. Rounding can merge distinct distances or reverse their order, and intermediate values can overflow even when the final distance is representable. Similar robustness problems have long been studied in computational geometry~\cite{richard1997adaptive,kettner2008classroom}. For example, in a test, Faiss~\cite{11202651} returns no valid neighbors from a database containing 10 exact copies of the query, all at distance $0$. Indeed, Faiss itself acknowledges that even brute-force search can suffer from floating-point rounding errors\footnote{\url{https://github.com/facebookresearch/faiss/wiki/FAQ}}. In practice, not every application requires mathematically exact results, and floating-point errors may have little impact in routine use. We argue, however, that users who rely on exact answers need to be aware of when such errors can change the returned neighbors. Developers also need a systematic, reproducible benchmark to assess the numerical correctness of their implementations.

In response to these issues, we present Errpedia, a purpose-built dataset dedicated to systematically testing numerical correctness in exact vector search. Its 1,616 queries span 21 error categories to help discover a wide range of numerical errors. To the best of our knowledge, Errpedia is the first dataset of this kind.
\section{Sources of Error}
\label{sec:numerical-errors}

\noindent\textbf{Problem definition.}
Let $X=\{x_1,\ldots,x_n\}$ be a collection of $d$-dimensional vectors and $q$ a query vector in the same space. Each member has a unique identifier. Distinct members may have identical coordinates. All coordinates are finite IEEE 754 32-bit float values interpreted as exact reals. We use squared Euclidean distance $D(q,x)=\sum_{j=1}^{d}(q_j-x_j)^2$. Unless otherwise stated, distance refers to this measure. For $1\leq k\leq n$, $R_k(q)$ contains the $k$ nearest members of $X$, with identifiers breaking exact ties. Errpedia uses $k<n$ and a strictly positive top-$k$ boundary gap, making membership in $R_k(q)$ independent of tie-breaking. We independently verify each ground-truth set using exact rational arithmetic. We consider an unordered result set that contains exactly the identifiers of these $k$ members as correct.

Under this setting, even exhaustive search can violate this criterion when numerical errors alter comparisons or candidate selection for multiple reasons.

\noindent\textbf{Rounding and accumulation.}
Evaluating $D(q,x)$ in floating point produces a computed distance score $\widehat D(q,x)$ through subtraction, multiplication, and addition, each of which can round its exact result to the working precision~\cite{goldberg1991floating}. Small terms may be absorbed into a large partial sum, and changing the accumulation order can change the result. Distinct distances can then become tied or reversed. Errors comparable to the gap at the top-$k$ boundary can change the returned set. Wider arithmetic can reduce errors but does not certify the order.

\noindent\textbf{Cancellation.}
An equivalent expression is
\begin{equation}
  D(q,x)=\|q\|^2+\|x\|^2-2q^\top x.
  \label{eq:norm-dot-distance}
\end{equation}
For nearby vectors with large norms, this expression subtracts large, already rounded quantities to obtain a small residual. Errors in those quantities can dominate the true distance and obscure the ordering of candidates. Algebraic equivalence therefore does not guarantee equivalent numerical behavior.

\noindent\textbf{Overflow and underflow.}
Finite coordinates do not ensure that every intermediate result is representable. A difference, square, or partial sum can overflow to infinity, while a tiny nonzero term can round to $0$. Subsequent operations involving infinities can produce NaN. Intermediates can be out of range even when the final distance is representable, so checking inputs alone cannot rule out range failures.

\noindent\textbf{Score conversion.}
Distance evaluation may be followed by a square root, conversion to a similarity score, or a reduction in precision. Strictly monotone transformations retain distance distinctions in real arithmetic, but their floating-point implementations can map different inputs to the same output. Selecting the top-$k$ by these scores can lose distinctions preserved by the distance computation and select a farther candidate from a computed tie.

\noindent\textbf{Candidate handling.}
Top-$k$ collection must distinguish valid candidates from unused slots. Using an extreme distance score for an empty slot can make it indistinguishable from a candidate with that score. If insertion requires strict improvement, the candidate may be rejected while a slot is still empty. NaN distance scores also require explicit handling because ordered comparisons with them are false. These interactions can leave fewer than $k$ neighbors despite enough eligible vectors.

\noindent\textbf{Other sources.}
Additional failures can arise from unintended coordinate changes during parsing or type conversion, nonconservative pruning bounds that discard true neighbors, or output and transport failures that prevent complete results from reaching the caller. These mechanisms can interact: overflow during distance computation may become missing neighbors when combined with candidate handling.
\begingroup
\dimen\footins=100pt
\begin{savenotes}
\section{The Errpedia Dataset}
\label{sec:dataset}

\subsection{Specification}
\label{sec:dataset-specification}

Errpedia contains a vector collection $X$ with $n=22{,}501$ members and a query collection $Q=(q_1,\ldots,q_b)$ with $b=1{,}616$, $d=128$, and $k=10$. All coordinates are finite 32-bit float values. Each query $q\in Q$ defines a test case over $X$ with a unique ground-truth top-10 result set as defined in Section~\ref{sec:numerical-errors}. The 21 categories form 6 evaluation groups (Table~\ref{tab:dataset-groups}).

\begin{table}[t]
  \centering
  \footnotesize
  \setlength{\tabcolsep}{3pt}
  \caption{Evaluation groups in Errpedia.}
  \label{tab:dataset-groups}
  \begin{tabular}{@{}lrr@{}}
    \toprule
    Group (code) & Categories & Queries \\
    \midrule
    Arithmetic precision (AP) & 7 & 1,263 \\
    Range limits (RL) & 5 & 109 \\
    Norm-dot-product evaluation (ND) & 2 & 69 \\
    16-bit float computation (FP) & 4 & 125 \\
    Similarity mappings (SM) & 2 & 38 \\
    Input representation (IR) & 1 & 12 \\
    \midrule
    Total & 21 & 1,616 \\
    \bottomrule
  \end{tabular}
\end{table}

\begin{table}[t]
  \centering
  \small
  \caption{Query counts by category.}
  \label{tab:dataset-categories}
  \begin{tabular}{@{}p{0.42\columnwidth}@{\hspace{0.04\columnwidth}}p{0.54\columnwidth}@{}}
    \toprule
    Category \hfill Queries & Category \hfill Queries \tabularnewline
    \midrule
    \raggedright\emph{Arithmetic precision (AP)} &
    \raggedright\emph{\mbox{Norm-dot-product evaluation (ND)}} \tabularnewline
    \texttt{sub} \hfill 58 & \texttt{normdot\_cancel} \hfill 49 \tabularnewline
    \texttt{sq\_round} \hfill 59 & \texttt{normdot\_overflow} \hfill 20 \tabularnewline
    \texttt{absorb} \hfill 150 & \raggedright\emph{Similarity mappings (SM)} \tabularnewline
    \texttt{order\_disc} \hfill 480 & \texttt{sim\_saturation} \hfill 30 \tabularnewline
    \texttt{neartie} \hfill 405 & \texttt{exp\_similarity} \hfill 8 \tabularnewline
    \texttt{collision\_ladder} \hfill 83 & \raggedright\emph{Input representation (IR)} \tabularnewline
    \texttt{sqrt} \hfill 28 & \texttt{decimal\_ingest} \hfill 12 \tabularnewline
    \raggedright\emph{Range limits (RL)} & \raggedright\emph{16-bit float computation (FP)} \tabularnewline
    \texttt{sq\_underflow} \hfill 50 & \texttt{fp16\_term} \hfill 14 \tabularnewline
    \texttt{sq\_subnormal} \hfill 28 & \texttt{fp16\_underflow} \hfill 96 \tabularnewline
    \texttt{overflow\_inf} \hfill 19 & \texttt{fp16\_absorb} \hfill 6 \tabularnewline
    \texttt{diff\_overflow} \hfill 6 & \texttt{fp16\_overflow} \hfill 9 \tabularnewline
    \texttt{max\_sentinel} \hfill 6 & \tabularnewline
    \bottomrule
  \end{tabular}
\end{table}

\noindent\textbf{Contents.}
The release stores $X$ in \path{data/base.csv}, 1 vector per row without a header. Identifiers are 0-based row indices. Query JSON files occupy 6 group directories under \path{data/}: \texttt{ap}, \texttt{rl}, \texttt{nd}, \texttt{fp}, \texttt{sm}, and \texttt{ir}. For example, \path{data/ap/sub_01.json} stores \texttt{case\_id}, \texttt{category}, \texttt{query\_row} (the original 0-based query index), \texttt{query} (the coordinates), and \texttt{gt\_ids} (the 10 ground-truth identifiers). Coordinates use complete exact decimal values, preserving the inputs under correctly rounded 32- or 64-bit float parsing.

\noindent\textbf{Applicability.}
Some systems impose numerical constraints on their inputs. We mark a query as not applicable when its inputs do not satisfy these constraints. Although an implementation may accept such inputs and even return correct results, evaluation should respect the input requirements specified by its developers.

\noindent\textbf{Scoring.}
A query passes only if it returns exactly the 10 distinct ground-truth identifiers, regardless of order or rounding in reported distances. For an evaluation group $G$, let $\mathcal{C}_G$ contain its categories with applicable, executed queries, and let $p_c$ be the number of these $n_c$ queries that pass in category $c$. The group score is
\begin{equation}
  \operatorname{Score}(G)=\frac{100}{|\mathcal{C}_G|}
  \sum_{c\in\mathcal{C}_G}\frac{p_c}{n_c}.
  \label{eq:group-score}
\end{equation}
Categories therefore receive equal weight within a group despite different query counts. Overall is the mean pass rate across all applicable, executed categories. Report raw passed/total query counts alongside these scores. Categories with no applicable, executed queries are marked N/A. Input rejections and execution failures are reported separately.

\subsection{Construction}
\label{sec:dataset-construction}

\noindent\textbf{General Methodology.}
Each case starts from a small candidate set. A typical case contains 9 vectors closer to the query than 2 remaining candidates. These 2 candidates should occupy ranks 10 and 11, respectively. We choose their coordinates so that a floating-point model of the computation can lose the difference between their distances.

\noindent\textbf{Arithmetic precision (AP).}
The \texttt{sub} construction pairs a large query coordinate with small, opposite-sign candidate coordinates, making distinct exact differences round identically. For \texttt{sq\_round}, we search for 2-term squared sums whose order is lost through product rounding. The \texttt{absorb} construction adds a nonzero squared term smaller than $1/2$ of the local spacing of a large partial sum, while \texttt{order\_disc} varies term order and placement across accumulator lanes. To expose small gaps, \texttt{neartie} perturbs coordinate-permuted distance patterns. The \texttt{collision\_ladder} construction places several strictly ordered candidates within a region of tied distance scores in the floating-point model. The \texttt{sqrt} construction uses distinct squared distances whose rounded square roots coincide in the targeted precision.

\noindent\textbf{Range limits (RL).}
For \texttt{sq\_underflow}, we perturb exact matches so that nonzero squared differences round to $0$ or are cleared under flush-to-zero. The \texttt{sq\_subnormal} construction searches for ordered squared sums that tie or reverse after product rounding on the 32-bit float subnormal grid. In \texttt{overflow\_inf}, individual squares or their sum overflow. The \texttt{diff\_overflow} cases use large, finite, opposite-sign coordinates whose differences exceed the 32-bit float range. The \texttt{max\_sentinel} construction makes boundary distance scores round to the largest finite 32-bit float value, testing implementations that also use it as an empty-slot threshold.

\noindent\textbf{Norm-dot-product evaluation (ND).}
The \texttt{normdot\_cancel} construction uses identical or nearby vectors with large norms, so cancellation in the expanded distance formula obscures their small exact distances. The \texttt{normdot\_overflow} cases contain 10 exact query matches whose individual norm terms or accumulated norms overflow in the same formula.

\noindent\textbf{16-bit float computation (FP).}
These categories target IEEE 754 16-bit float operations on 32-bit float inputs. The \texttt{fp16\_term} construction uses coordinates representable in 16-bit float whose squared sums tie or reverse after product rounding. For \texttt{fp16\_underflow}, differences remain nonzero in 16-bit float but their squares round to $0$. The \texttt{fp16\_absorb} cases place small terms after a large term in 1 accumulator lane. The \texttt{fp16\_overflow} cases overflow 16-bit float products or lane sums while keeping inputs representable.

\noindent\textbf{Similarity mappings (SM).}
The \texttt{sim\_saturation} construction pairs $0$ and small positive squared distances $D$ whose 32-bit float similarity scores $1/(1+D)$ both become $1$. For \texttt{exp\_similarity}, distinct distances produce 32-bit float similarity scores $\exp(-\sqrt{D})$ that underflow to $0$.

\noindent\textbf{Input representation (IR).}
For \texttt{decimal\_ingest}, we search for distinct 32-bit float coordinates that serialize to identical 7-significant-digit text. The serialization erases the distinction at an exact-match boundary.

{\clubpenalty=10000
\noindent\textbf{Packing and ground truth.}
We pack the candidate sets into 1 shared base collection. Where coordinate positions are available, we give a query and its candidates shared values at those positions to separate them from other candidate sets. These values cancel in direct coordinate differences but still participate in norm-dot-product evaluation, so validation uses the complete vectors. Every finite 32-bit float coordinate is an integer multiple of $2^{-149}$. Squared distances can therefore be compared using exact integer arithmetic in units of $2^{-298}$. We check the ground-truth sets against the shared collection and independently verify them using exact rational arithmetic to ensure that other candidate sets do not change a query's top-10 result set.
\par}

\noindent\textbf{Coverage and selection.}
Constructions vary target precision, scale, coordinate position, accumulator layout, candidate count, and identifier placement. We generate cases by formula and search, retaining only those with a strictly positive top-$k$ boundary gap lost in the targeted floating-point model.

\begin{table*}[!t]
  \centering
  \small
  \setlength{\tabcolsep}{3pt}
  \renewcommand{\arraystretch}{0.95}
  \caption{Evaluation results for 20 configurations. Bold indicates the highest pass rate(s) in each group.}
  \label{tab:evaluation-results}
  \fontsize{9}{10.8}\selectfont
  \begin{tabular*}{\textwidth}{@{\extracolsep{\fill}}llrrrrrrr@{}}
    \toprule
    System & Type & \multicolumn{1}{c}{AP} & \multicolumn{1}{c}{RL} & \multicolumn{1}{c}{ND} & \multicolumn{1}{c}{FP} & \multicolumn{1}{c}{SM} & \multicolumn{1}{c}{IR} & \multicolumn{1}{c}{Overall\textsuperscript{*}} \\
    \midrule
    Faiss v1.15.1\footnote{\url{https://faiss.ai/}} & Vector library & 572/1263 & 41/109 & 24/69 & 94/125 & \textbf{38/38} & 6/12 & 49.0\% \\
    scikit-learn v1.9.1\footnote{\url{https://scikit-learn.org/stable/}} & Machine learning library & 1181/1263 & 102/109 & 43/69 & \textbf{125/125} & \textbf{38/38} & 6/12 & 86.5\% \\
    SciPy v1.18.1\footnote{\url{https://scipy.org/}} & Scientific library & \textbf{1189/1263} & \textbf{109/109} & \textbf{69/69} & \textbf{125/125} & \textbf{38/38} & \textbf{12/12} & \textbf{95.6\%} \\
    \addlinespace[1pt]
    Milvus v3.0.5\footnote{\url{https://milvus.io/}} & Vector database & 829/1263 & 40/109 & 68/69 & \textbf{125/125} & \textbf{38/38} & \textbf{12/12} & 64.4\% \\
    Qdrant v1.20.1\footnote{\url{https://qdrant.tech/}} & Vector database & 780/1263 & 27/109 & 66/69 & \textbf{125/125} & \textbf{38/38} & \textbf{12/12} & 62.7\% \\
    Weaviate v1.40.5\footnote{\url{https://weaviate.io/}} & Vector database & 928/1263 & 54/109 & 68/69 & \textbf{125/125} & \textbf{38/38} & \textbf{12/12} & 73.8\% \\
    LanceDB v0.41.0\footnote{\url{https://www.lancedb.com/}} & Vector database & 767/1263 & 44/109 & 68/69 & \textbf{125/125} & \textbf{38/38} & \textbf{12/12} & 66.4\% \\
    \addlinespace[1pt]
    ClickHouse v26.8.34.13\footnote{\url{https://clickhouse.com/}} & SQL database & 798/1263 & 34/109 & 65/69 & \textbf{125/125} & \textbf{38/38} & \textbf{12/12} & 66.3\% \\
    DuckDB v2.0.0\footnote{\url{https://duckdb.org/}} & SQL database & 943/1263 & 56/109 & 68/69 & \textbf{125/125} & \textbf{38/38} & \textbf{12/12} & 75.4\% \\
    TiDB v8.5.8\footnote{\url{https://www.pingcap.com/tidb/}} & SQL database & 870/1263 & 57/109 & 68/69 & \textbf{125/125} & \textbf{38/38} & \textbf{12/12} & 73.7\% \\
    MariaDB v12.3.3\footnote{\url{https://mariadb.org/}}\textsuperscript{,$\dagger$} & SQL database & 1172/1263 & \textbf{78/78} & \textbf{49/49} & \textbf{125/125} & \textbf{38/38} & \textbf{12/12} & 92.5\% \\
    pgvector v0.8.7\footnote{\url{https://github.com/pgvector/pgvector}} & SQL extension & 958/1263 & 56/109 & 68/69 & \textbf{125/125} & \textbf{38/38} & \textbf{12/12} & 74.8\% \\
    \addlinespace[1pt]
    Elasticsearch v9.5.6\footnote{\url{https://www.elastic.co/elasticsearch}}\textsuperscript{,$\dagger$} & Search engine & 648/1263 & 0/78 & 38/49 & \textbf{125/125} & 13/38 & 8/12 & 55.8\% \\
    OpenSearch v3.9.0\footnote{\url{https://opensearch.org/}} & Search engine & 616/1263 & 4/109 & 56/69 & 86/125 & 8/38 & 8/12 & 48.8\% \\
    Apache Solr v10.1.0\footnote{\url{https://solr.apache.org/}} & Search engine & 592/1263 & 4/109 & 56/69 & 86/125 & 8/38 & 8/12 & 48.5\% \\
    \addlinespace[1pt]
    Redis v7.4.7\footnote{\url{https://redis.io/}} & Key-value database & 829/1263 & 44/109 & 68/69 & \textbf{125/125} & \textbf{38/38} & \textbf{12/12} & 67.6\% \\
    Dragonfly v2.1\footnote{\url{https://www.dragonflydb.io/}} & Key-value database & 774/1263 & 44/109 & 68/69 & \textbf{125/125} & \textbf{38/38} & \textbf{12/12} & 66.3\% \\
    \addlinespace[1pt]
    MongoDB v9.0.8\footnote{\url{https://www.mongodb.com/}} & Document database & 584/1263 & 2/109 & 47/69 & 64/125 & 8/38 & 7/12 & 43.6\% \\
    Neo4j v5.26.12\footnote{\url{https://neo4j.com/}} & Graph database & 770/1263 & 6/109 & 56/69 & 86/125 & 8/38 & 8/12 & 54.2\% \\
    SurrealDB v4.0.1\footnote{\url{https://surrealdb.com/}} & Multi-model database & 1163/1263 & \textbf{109/109} & \textbf{69/69} & \textbf{125/125} & \textbf{38/38} & \textbf{12/12} & 93.1\% \\
    \bottomrule
  \end{tabular*}
  \par\smallskip
  \begin{minipage}{\textwidth}
    \footnotesize
    \textsuperscript{*} Overall averages category pass rates with equal weight.
    \par\noindent $\dagger$ Inapplicable queries are excluded.
  \end{minipage}
\end{table*}

The compact release preserves declared coverage of construction variants, parameter ranges, candidate identifier orders, floating-point model behavior, and numerical endpoints. Selection is independent of measured product outcomes. Category sizes reflect the configurations needed to meet these requirements. For example, the 480 \texttt{order\_disc} queries conservatively retain strata for the number and placement of small terms under both candidate identifier orders. The 6 \texttt{max\_sentinel} queries cover 1 threshold regime, model behavior, and numerical endpoints. Different parameter settings can produce the same model outcome.
\section{Evaluation}
\label{sec:evaluation}

We evaluate 20 libraries and database systems in exact-search or exhaustive-search configurations, with $k=10$. We check each configuration's input conditions and compare its returned neighbor identifiers with the ground-truth sets.

18 configurations search all 22,501 base vectors for all 1,616 queries. Elasticsearch and MariaDB each reject 578 base vectors at ingestion. For these 2 configurations, we search the 21,923 admitted vectors and score the 1,565 applicable queries whose original ground-truth sets remain intact. We retain the remaining 51 queries in the dataset and mark them not applicable for these configurations.

\FloatBarrier
\end{savenotes}
\endgroup

Table~\ref{tab:evaluation-results} reports passed/total query counts by system type and evaluation group, together with the Overall score. All 20 configurations return mathematically incorrect top-$k$ result sets on AP queries within their scored scopes. RL and ND also expose failures in 17 configurations each. FP and SM each expose failures in 5 configurations, and IR exposes failures in 7.

The failure profiles differ substantially across systems. Among the 18 configurations tested on the complete collection, SciPy and SurrealDB fail 74 and 100 queries, respectively, all in AP. Both pass every query in the other 5 groups. OpenSearch, Apache Solr, MongoDB, and Neo4j each fail queries in all 6 groups. The differences are especially clear in RL: scikit-learn passes 102/109 queries, OpenSearch and Apache Solr each pass 4/109, and MongoDB passes 2/109. Elasticsearch also fails all 78 applicable RL queries in its admitted-row collection. MariaDB fails 91 of its 1,565 applicable queries, all in AP.
\section{Failure Analysis}
\label{sec:failure-analysis}

We analyze 1 representative case from each of the 6 groups to explain how numerical errors lead to incorrect search results. For each case, we trace how the error arises, show its effect on the returned neighbors, and discuss how to prevent it.

\subsection{Arithmetic precision (AP)}

Case~\texttt{absorb\_129} exposes a distinction lost even in 64-bit floats. Its boundary candidates have exact squared distances $R^2$ and $R^2+t^2$, where $R=41552892$ and $t=265/8192$. The positive gap $t^2\approx 0.001046$ is smaller than $1/2$ of the 64-bit float spacing $0.25$ at $R^2$. SciPy~v1.18.1's 4-accumulator distance kernel\footnote{\url{https://github.com/scipy/scipy/blob/e4e854eaa8f18d807cd3496028e257e36caa93cc/scipy/spatial/ckdtree/src/distance.h\#L55-L82}} therefore produces the same distance score $\widehat D(q,x)$ for both candidates. Distance probes also return the same unsquared Euclidean distance, $R$. The full-collection search selects the farther candidate at the boundary.

The computation has lost the information needed to distinguish the candidates. Changing the rule for breaking computed ties cannot guarantee the correct set. An uncertain boundary comparison needs more accurate evaluation, ultimately exact arithmetic if the distinction remains unresolved.

\subsection{Range limits (RL)}

Case~\texttt{max\_sentinel\_02} makes Milvus~v3.0.5 return 9 neighbors for $k=10$. 9 local candidates have distance scores below \texttt{FLT\_MAX}. The remaining 2 boundary distance scores round to that finite value. The underlying Knowhere collector initializes empty slots to \texttt{FLT\_MAX} and accepts candidates only when their distance scores strictly improve the threshold\footnote{\url{https://github.com/zilliztech/knowhere/blob/db9598469caba6b2d9b85406ecda3b3c6edd5d9a/thirdparty/faiss/faiss/impl/ResultHandler.h\#L273-L365}. \url{https://github.com/zilliztech/knowhere/blob/db9598469caba6b2d9b85406ecda3b3c6edd5d9a/thirdparty/faiss/faiss/utils/ordered_key_value.h\#L64-L79}}. Neither boundary candidate can therefore fill the 10th slot.

Collectors should track occupancy independently of distance scores and admit eligible candidates while slots remain empty. This restores completeness for the finite sentinel collision. Resolving the rounded distance tie still requires an accurate boundary comparison.

\subsection{Norm-dot-product evaluation (ND)}

Case~\texttt{normdot\_overflow\_17} contains 10 exact copies of the query. Each individual coordinate square is finite in 32-bit float, but their norm sum overflows. Faiss~v1.15.1's BLAS path\footnote{\url{https://github.com/facebookresearch/faiss/blob/75c755c5a1fc854756aaf3bd1131213aeb21d8d1/faiss/utils/distances.cpp\#L443-L501}} evaluates Eq.~\ref{eq:norm-dot-distance}, producing an indeterminate expression involving infinities even for an exact match. The full-collection search returns 10 invalid identifier slots, omitting every neighbor at distance $0$. A local test uses the 11-vector candidate set and a fixed batch of 1,616 query copies. Changing only the dispatch threshold to select direct differences restores all 10 matches with distances of $0$.

Direct differences avoid this failure for identical vectors. More generally, nonfinite intermediate results require recomputation with sufficient range. Candidate selection must not silently discard a possible neighbor because its provisional distance is NaN.

\subsection{16-bit float computation (FP)}

Case~\texttt{fp16\_underflow\_13} isolates product underflow in a 16-bit float model. The inputs are exactly representable in 16-bit float. A nonmatching vector differs from 10 exact query copies by $2^{-13}$ in 1 coordinate. Its squared distance is $2^{-26}$, below $1/2$ of the smallest positive 16-bit float subnormal, $2^{-24}$. The product rounds to $0$, creating a computed tie. The model then includes the nonmatch among its 10 results.

Promoting the operands before multiplication preserves this term in 32-bit float and restores the local top-10 result set. Widening only the accumulator cannot recover a product already rounded to $0$. The model isolates underflow during multiplication, before accumulation.

\subsection{Similarity mappings (SM)}

Case~\texttt{sim\_saturation\_04} has 10 exact query copies and a perturbed candidate with $D=9\cdot 2^{-56}>0$. This distance survives 32-bit float evaluation, but $1+D$ rounds to $1$. The Lucene~v9.11.1 mapping $1/(1+D)$ used by Neo4j~v5.26.12\footnote{\url{https://github.com/apache/lucene/blob/0c087dfdd10e0f6f3f6faecc6af4415e671a9e69/lucene/core/src/java/org/apache/lucene/index/VectorSimilarityFunction.java\#L32-L42}} therefore assigns it the same similarity score as a match at distance $0$. In the full-collection search, all 10 returned identifiers lie outside the exact-match set, and all have reported similarity score $1$. The mapping hides distinct exact distances.

To preserve distance order, selection should retain the unmapped distance and refine uncertain comparisons. Similarity scores may be rounded after the neighbor set has been selected. Reranking only the returned identifiers cannot recover matches already excluded by score saturation.

\subsection{Input representation (IR)}

Case~\texttt{decimal\_ingest\_05} illustrates information loss before distance evaluation. Its decisive coordinates are 12570212 and 12570213, both exact 32-bit float values. In a serialization test, a serializer retaining 7 significant decimal digits writes both as \texttt{1.257021e+07}. Parsing this text merges an exact match and a nonmatch, and a search over the parsed vectors returns an incorrect top-10 result set.

Base and query vectors should be transported in binary or decimal text that preserves every 32-bit float value. 9 significant digits suffice with correctly rounded 32-bit float parsing. Our 9-digit serialization test preserves every local coordinate and restores the correct top-10 result set. Wider arithmetic after 7-digit ingestion cannot recover the original distinction.

\subsection{Protecting True Neighbors}
The fixes above address specific errors. To guarantee correct results, a system must also avoid discarding true neighbors during search. If floating-point calculations cannot reliably determine which of 2 candidates is closer, keep both until an exact comparison resolves their order~\cite{richard1997adaptive}. A tree branch should likewise be skipped only when its vectors are certainly too far away. Recomputing distances for the final $k$ results cannot recover a true neighbor discarded earlier.

\section{Conclusion}
\label{sec:conclusion}

We present Errpedia, a dataset for testing the numerical correctness of exact vector search. It contains 1,616 queries across 21 categories, with unique ground-truth sets verified using exact rational arithmetic. Each of the 20 tested configurations returns mathematically incorrect top-$k$ result sets for some applicable queries. Errpedia provides reproducible cases for validating implementations, checking benchmark ground truth, and detecting numerical regressions.

Errpedia has 2 main known limitations: it covers only squared Euclidean distance and uses fixed parameters of $d=128$ and $k=10$. Extending its coverage to other distance measures and parameter settings remains future work.

\flushcolsend
\FloatBarrier
\bibliographystyle{plain}
\bibliography{ref}
\end{document}